\documentclass[journal,10pt,twocolumn]{IEEEtran}

\usepackage{amsmath,amssymb}
\usepackage{bbm}
\usepackage{graphicx}
\graphicspath{{figures/}}
\usepackage{booktabs}
\usepackage{multirow}
\usepackage{array}
\usepackage[colorlinks=true,linkcolor=blue,citecolor=blue,urlcolor=blue]{hyperref}
\usepackage[numbers,sort&compress]{natbib}
\usepackage{microtype}
\usepackage{caption}
\usepackage{tikz}
\usetikzlibrary{arrows.meta, positioning, calc, fit}

\newcommand{\Kq}{K_{\mathrm{Q}}}
\newcommand{\Kc}{K_{\mathrm{C}}}
\newcommand{\ggeo}{g}
\newcommand{\vis}{V}
\newcommand{\Nph}{N}
\newcommand{\Mmod}{M}
\newcommand{\Dout}{D}
\newcommand{\alignK}{\mathcal{A}}
\usepackage{float}
\newcommand{\identity}{\mathbbm{1}}
\newcommand{\Kx}{K_{\mathrm{X}}}

\title{When Does Interference Help Learning? Kernel Geometry as a Pre-Experimental Test for Photonic Reservoir Computing}

\author{Oishik~Kar$^{1,2}$ and Aswath~Babu~H$^{1,3}$\\[2pt]
{\small $^{1}$Center of Excellence in Quantum AI and Computing, Indian Institute
of Information Technology Dharwad, Karnataka, India}\\[1pt]
{\small $^{2}$Department of Computer Science and Engineering, Indian
Institute of Information Technology Dharwad, Karnataka, India}\\
{\small $^{3}$Department of Arts, Science and Design, Indian Institute
of Information Technology Dharwad, Karnataka, India}\\[1pt]

{\small \texttt{23bcs089@iiitdwd.ac.in}\quad \texttt{aswath@iiitdwd.ac.in}%
\quad (corresponding author: Oishik Kar)}}

\begin{document}

\maketitle

% ================================================================
\begin{abstract}
Multi-photon interference is the resource most native to photonic quantum machine learning, yet experiments on photonic reservoirs have reported advantages, null results, and train-only effects. These outcomes cannot be compared: each measures task-dependent accuracy on different tasks and platforms. Here we resolve the question task-independently through the kernel geometry of boson-sampling feature maps. Interference redistributes feature variance across roughly twice as many effective dimensions as distinguishable-particle statistics, generating a large geometric separation between the corresponding kernels. The separation grows superlinearly with indistinguishability, is linear in Hong--Ou--Mandel visibility at small visibility for two to four photons, and retains approximately 80\% of its magnitude at the source quality of current quantum processors. Under experimental sampling budgets the separation is learnable on adversarially constructed tasks (accuracy advantage $+0.22 \pm 0.05$) and an order of magnitude smaller on natural tasks, gated by kernel--task alignment. The results reconcile the existing experimental record and supply a pre-experimental protocol: kernel geometry establishes that an interference-specific learning resource exists, and task alignment determines whether a given task can access it; both are computable before committing significant hardware resources.
\end{abstract}

% ================================================================
\section{Introduction}

Quantum reservoir computing replaces the trained dynamical core of a reservoir computer~\cite{jaeger2001,lukosevicus2009,tanaka2019} with a fixed quantum system~\cite{fujii2017,ghosh2019,nakajima2020,mujal2023,martinezpena2021,bravo2022,kornjaca2024,martinezpena2023}; photonics is a natural substrate: multi-photon interference in a linear interferometer
generates feature maps of dimension $\binom{\Mmod}{\Nph}$ from compact
circuits, with no trainable parameters in the quantum
layer~\cite{sakurai2025,rambach2026,sharifian2025,gong2025}, a paradigm
realized experimentally on photonic and cloud quantum
processors~\cite{suprano2024,delorenzis2025}. The interference
itself, however, is not what makes these maps expressive.
Distinguishable-photon statistics, the output of a classical
multi-particle process through the same interferometer, remain
high-degree nonlinear functions of the circuit parameters and retain
single-particle coherence~\cite{tichy2014,renema2018}. Whether interference
contributes anything to learning is therefore an empirical question, not
a foregone conclusion.

The experimental record is mixed. Time-series forecasting with two-photon
states in a reconfigurable integrated circuit found a clear advantage for
indistinguishable photons on nonlinear temporal tasks (temporal parity,
NARMA, and Mackey--Glass forecasting), while linear memory capacity was
indistinguishable between the two input regimes~\cite{dibartolo2026}; a
fibre-based quantum extreme learning machine could resolve no
distinguishability effect at two photons, while measuring a higher
feature-space rank for indistinguishable photons~\cite{joly2025}; and a
large-scale study validated on a photonic quantum processor reported that indistinguishability improves training accuracy while leaving test accuracy essentially unchanged~\cite{rambach2026}. These
outcomes are not necessarily contradictory; they are mutually
incomparable. Each reports task-dependent accuracies on different tasks,
readouts, and platforms, and task-dependent metrics cannot isolate an
effect that is modulated by the alignment between task and feature map.

Resolving this requires a task-independent characterization of the feature
maps themselves. Because a trained readout is a regularized linear model
on the reservoir output, its learning behavior is governed by the kernel
induced by the output features~\cite{schuld2021}. Kernel geometry can be
computed from the data and the circuit before any task is run, and it
bounds the performance of any linear readout on any task~\cite{huang2021,liu2021}. While the necessity of kernel geometry for quantum advantage is established in general~\cite{huang2021}, its magnitude and behavior under partial distinguishability in the reservoir setting, and its sufficiency in practice for natural tasks, have not been characterized.

Here we characterize the kernel geometry of boson-sampling reservoirs
across the full distinguishability dial, and show that this geometry
determines and bounds when interference aids learning. Multi-photon
interference redistributes feature variance across roughly twice as many
effective dimensions as distinguishable-particle statistics, producing a
geometric separation between the corresponding kernels that turns on
superlinearly with indistinguishability and survives hardware-typical
source imperfection and experimental sampling budgets. Whether this
separation is learnable, however, is governed not by the geometry but by
the alignment between kernel and task. We measure this alignment for
adversarially designed tasks and for natural tasks, and find a gated
relationship: a designed-task accuracy advantage
of $+0.222 \pm 0.046$, an order of magnitude smaller on natural tasks,
and zero where neither kernel aligns with the task. This
account reconciles the apparent disagreement among existing
experiments~\cite{joly2025,dibartolo2026,rambach2026,yin2025}, and yields
a practical protocol: the kernel geometry and the task alignment can be
computed before committing significant hardware resources.

\begin{figure*}[htbp]
\centering
\includegraphics[width=0.82\linewidth]{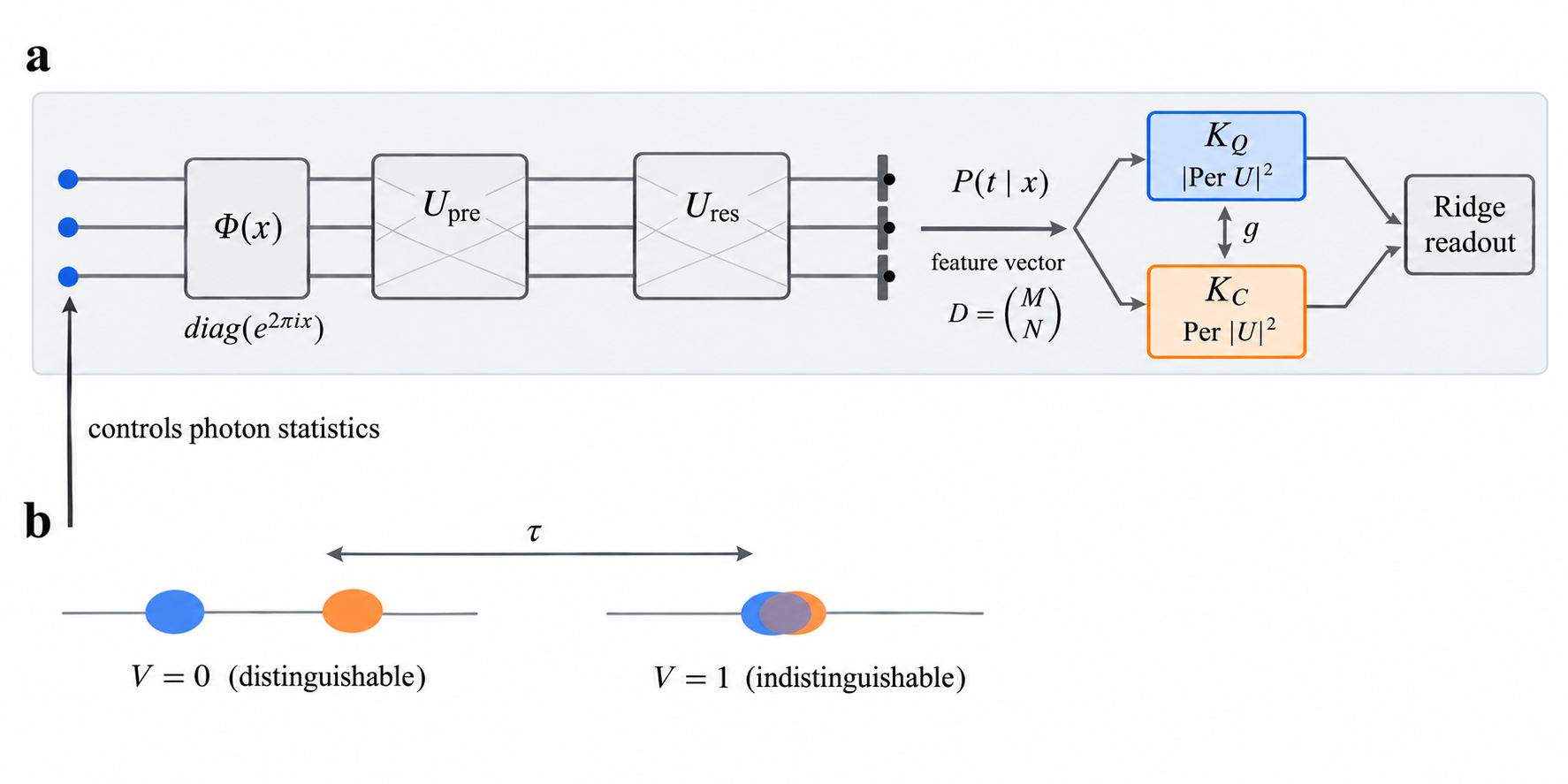}
\caption{\textbf{Boson-sampling reservoir kernel estimation and the
distinguishability dial.} \textbf{a}, An input $x \in [0,1]^{d}$ is
encoded as phase shifts between two Haar-random interferometers;
 $\Nph$ injected photons, post-selected on coincidences, yield the
outcome distribution $P(t \mid x)$ of dimension
 $\binom{\Mmod}{\Nph}$. Indistinguishable and distinguishable photons
generate the kernels $\Kq$ and $\Kc$, whose geometric difference
 $\ggeo$ (Eq.~(6)) quantifies the separation. \textbf{b}, The
Hong--Ou--Mandel visibility $\vis = I^{2}$ controls the photon
statistics; this is the dial swept in Fig.~2.}
\label{fig:arch}
\end{figure*}

% ================================================================
\section{Boson-Sampling Reservoirs and Their Kernels}
\label{sec:setup}

We consider the standard photonic reservoir architecture (Fig. 1)~\cite{sakurai2025,rambach2026,cimini2025}. A classical input
$x \in [0,1]^{d}$ is encoded into the phases of a linear interferometer,
\begin{equation}
\begin{split}
U(x) \;&=\; U_{\mathrm{res}}\,\Phi(x)\, U_{\mathrm{pre}}, \\
\Phi(x) \;&=\; \mathrm{diag}\!\left(e^{2\pi i x_1},\dots,
e^{2\pi i x_d},1,\dots,1\right),
\end{split}
\label{eq:encoding}
\end{equation}
where $U_{\mathrm{pre}}$ and $U_{\mathrm{res}}$ are independent Haar-random
$\Mmod \times \Mmod$ interferometers~\cite{clements2016,reck1994}. $\Nph$ photons are
injected, one per mode, into the first $\Nph$ input modes. Photon-number
measurements at the output, post-selected on $\Nph$ detection events and
renormalized, yield a feature vector of dimension
$\Dout = \binom{\Mmod}{\Nph}$: for each collision-free outcome
$t$ (a set of $\Nph$ distinct output modes),
\begin{align}
P_{\mathrm{Q}}(t \mid x) \;&=\;
\left| \operatorname{Per} U(x)[s,t] \right|^{2},
\label{eq:PQ}\\
P_{\mathrm{C}}(t \mid x) \;&=\;
\operatorname{Per} \left| U(x)[s,t] \right|^{2},
\label{eq:PC}
\end{align}
where $s$ denotes the occupied input modes,
$U(x)[s,t]$ is the $\Nph \times \Nph$ submatrix with rows $s$ and
columns $t$, and $\operatorname{Per}$ is the matrix
permanent~\cite{aaronson2011,gurvits2011,clifford2018}. The two statistics differ only in the
placement of the modulus: Eq.~\eqref{eq:PQ} describes indistinguishable
photons, whose amplitudes interfere; Eq.~\eqref{eq:PC} describes
distinguishable photons, a classical multi-particle process. Both are
degree-$2\Nph$ nonlinear functions of the input phases, so any difference
in learning behavior reflects the sign structure that interference
introduces, not the order of the nonlinearity.

Partial distinguishability is controlled by the mutual mode overlaps of
the injected photons. Writing the $\Nph \times \Nph$ overlap matrix
$S$ with entries $S_{ij} = \langle \varphi_i \mid \varphi_j \rangle$
between the internal states of photons $i$ and $j$, the outcome
probabilities become~\cite{tichy2014,shchesnovich2022,nerenberg2024,renema2018}
\begin{equation}
\begin{split}
P(t \mid x) \;=\;&
\frac{1}{\prod_k t_k!\,\prod_j s_j!}
\sum_{\sigma,\tau \in \mathcal{S}_\Nph}
\!\!\Big( \prod_{k=1}^{\Nph} S_{\sigma(k),\tau(k)} \Big) \\
&\times \prod_{k=1}^{\Nph}
U(x)_{t_k,\,s_{\sigma(k)}}\,
\overline{U(x)_{t_k,\,s_{\tau(k)}}},
\end{split}
\label{eq:partial}
\end{equation}
which reduces to Eqs.~\eqref{eq:PQ} and~\eqref{eq:PC} for
$S = \mathbf{1}$ and $S = \mathbbm{1}$, respectively. We parameterize
uniform partial distinguishability, $S_{ij} = I$ for $i \neq j$, and
report the dial in terms of the Hong--Ou--Mandel visibility~\cite{hong1987}
$\vis = I^{2}$, the convention of the experiments we compare
against~\cite{rambach2026,yin2025}.

Because the trained readout is ridge regression on the output features,
the reservoir is equivalent to a kernel machine~\cite{havlicek2019,schuld2019,schuld2021,sakurai2025}. We study the Gram matrices
\begin{equation}
\begin{split}
\big(\Kq\big)_{ij} &= \sum_{t} P_{\mathrm{Q}}(t \mid x_i)\,
P_{\mathrm{Q}}(t \mid x_j), \\
\big(\Kc\big)_{ij} &= \sum_{t} P_{\mathrm{C}}(t \mid x_i)\,
P_{\mathrm{C}}(t \mid x_j),
\end{split}
\label{eq:kernels}
\end{equation}
over a dataset $\{x_i\}_{i=1}^{n}$, with the analogous partially
distinguishable kernel $K(\vis)$ obtained from Eq.~\eqref{eq:partial}.
The raw kernels of Eq.~\eqref{eq:kernels} carry no preprocessing degrees
of freedom and are used for all geometric statements; the standardized
feature kernels used by the operational pipelines of
refs.~\cite{sakurai2025,rambach2026} give the same spectral contrast
(Supplementary Note~1).

% ================================================================
\section{Interference Broadens the Feature Spectrum}
\label{sec:mechanism}

\begin{figure}[htbp]
\centering
\includegraphics[width=\linewidth]{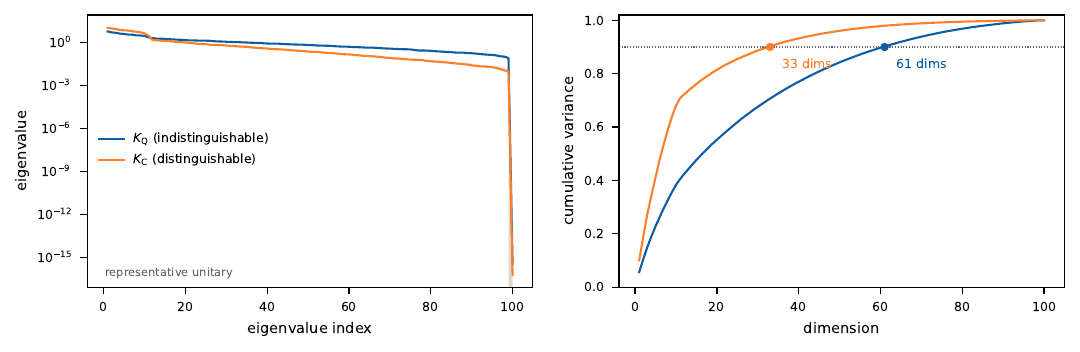}
\caption{\textbf{Multi-photon interference broadens the feature
spectrum.} \textbf{(a)} Sorted eigenvalues of the indistinguishable-
($\Kq$, blue) and distinguishable-photon ($\Kc$, orange) Gram matrices
over $n = 100$ inputs ($\Mmod = 12$, $\Nph = 3$, $d = 12$;
$\Dout = 220$). \textbf{(b)} Cumulative variance. In a representative
interferometer realization the leading eigenvalue of $\Kc$ exceeds that
of $\Kq$ (10.0 vs.\ 5.7), while $\Kq$ reaches 90\% of its trace in 61
dimensions against 33 for $\Kc$. Shaded regions: spread over 10 Haar
realizations.}
\label{fig:spectra}
\end{figure}

Fig.~\ref{fig:spectra} contrasts the two kernels at the flagship
configuration ($\Mmod = 12$, $\Nph = 3$, $d = 12$, $n = 100$). Both Gram
matrices have identical trace by construction, but they distribute it
differently: the distinguishable-photon kernel concentrates its variance
in a steep leading spectrum, reaching 90\% of its trace in 33 effective
dimensions, whereas the indistinguishable-photon kernel spreads variance
across 61, roughly twice as many. The contrast persists in the tails:
the smallest eigenvalue of $\Kc$ sits an order of magnitude below that
of $\Kq$.

The mechanism is the sign structure of the two statistics. Entries of
$P_{\mathrm{C}}$ are permanents of nonnegative matrices and cannot cancel;
fluctuations of the feature map across inputs therefore accumulate in a
low number of directions. Interference in $P_{\mathrm{Q}}$ introduces
signed amplitudes whose cancellations decorrelate the outcome
fluctuations, spreading the same total variance across many more
directions of the outcome space. This is the geometric signature behind
the expressivity arguments of refs.~\cite{joly2025,dibartolo2026}: Joly
et al.\ measure a higher feature-space rank for indistinguishable
photons in hardware at two photons, and Di Bartolo et al.\ attribute
their advantage to the richer transformations that indistinguishable
states afford.

Every geometric quantity studied below is downstream of this spectral contrast.

% ================================================================
\section{A Geometric Dial Controlled by Indistinguishability}
\label{sec:dial}

\begin{figure}[htbp]
\centering
\includegraphics[width=\linewidth]{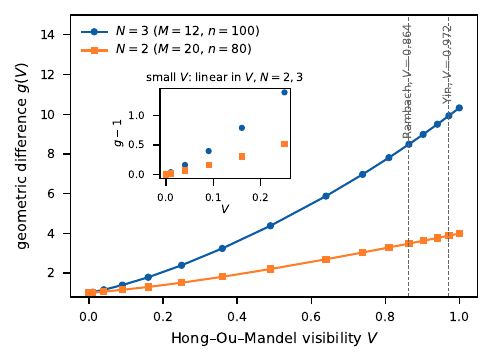}
\caption{\textbf{The geometric separation increases superlinearly with
visibility.} Geometric difference $\ggeo(\vis)$ between the
indistinguishable- and distinguishable-photon feature kernels
(square-root convention, Eq.~\eqref{eq:gdef}). Hardware source
qualities are marked: the quantum processor of ref.~\cite{rambach2026}
($\vis = 0.864$) and the interferometric kernel experiment of
ref.~\cite{yin2025} ($\vis = 0.972$). Inset: at small visibility,
$\ggeo - 1$ is linear in $\vis$ for $\Nph = 2, 3, 4$.}
\label{fig:dial}
\end{figure}

We quantify the separation by the geometric difference between the two
kernels~\cite{huang2021,yin2025},
\begin{equation}
\ggeo(\Kq, \Kc) \;=\;
\sqrt{\lambda_{\max}\!\left(
\Kq^{1/2}\,\big(\Kc + \varepsilon \mathbbm{1}\big)^{-1}\,
\Kq^{1/2} \right)},
\label{eq:gdef}
\end{equation}
with $\varepsilon$ a numerical regularization whose convergence we
document explicitly (Supplementary Note~1). The quantity $\ggeo^{2}$
upper-bounds the ratio of model complexities $s_{K}(y) = y^{\top} K y$
between the two kernels, so $\ggeo > 1$ certifies the existence of tasks
on which the distinguishable-photon readout is disadvantaged relative to
the indistinguishable one~\cite{huang2021}. The converse does not hold:
a large $\ggeo$ guarantees nothing about any particular task, a point to
which we return in Section~\ref{sec:gating}.

At full indistinguishability, the converged separation is
$\ggeo = 10.3 \pm 1.1$ over ten interferometer realizations
($\Mmod = 12$, $\Nph = 3$, $n = 100$). The dial between the two limits
is smooth and strongly nonlinear (Fig.~\ref{fig:dial}). At small
visibility, $\ggeo - 1$ is linear in $\vis$ for every photon number
examined: effective exponents
$k_{\vis} = 1.06, 1.09, 1.12$ for $\Nph = 2, 3, 4$ over the fitted windows
($\Nph = 4$ over a restricted small-visibility window). For $\Nph = 2$ this
linearity is exact: the mixed features are affine in $\vis$, so every kernel
entry is quadratic and first-order perturbation theory gives
$\ggeo(\vis) - 1 = (s/2)\,\vis + O(\vis^2)$ with $s = \lambda_{\max}$ of the
cross-statistics matrix (Supplementary Note~7, Proposition~1). The measured
exponents sit just above 1 for all $\Nph$, consistent with
pairwise-transposition dominance; the excess above unity is the convex
correction. Over the full dial, the effective exponent grows from 1.16 at
$\Nph = 2$ to 1.22 at $\Nph = 3$ (Supplementary Note~2).

Decoherence acts on this dial as a smooth attenuation rather than a
threshold. We define retention as the fraction of the excess separation
retained at visibility $\vis$,
\begin{equation}
R(\vis) \;=\; \frac{\ggeo(\vis) - 1}{\ggeo(1) - 1},
\label{eq:retention}
\end{equation}
and find $R = 80.3\% \pm 0.4\%$ at the source quality of the quantum
processor of ref.~\cite{rambach2026} ($\vis = 0.864$) and
$R = 95.8\% \pm 0.1\%$ at that of ref.~\cite{yin2025}
($\vis = 0.972$), both at $\Nph = 3$. Retention decreases with photon number: 83.1\% at $\Nph=2$ versus 80.4\% at $\Nph=3$ (both at $\vis=0.864$).

% ================================================================
\section{Robustness of the Separation}
\label{sec:robustness}

\begin{figure}[htbp]
\centering
\includegraphics[width=\linewidth]{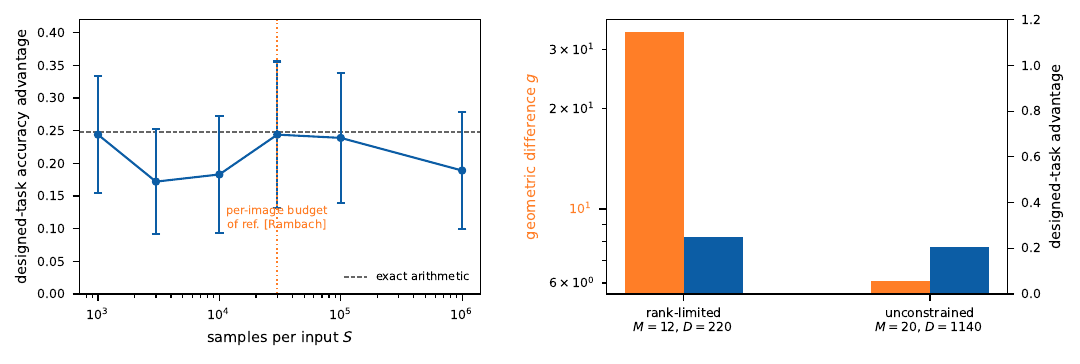}
\caption{\textbf{The separation is sampling-accessible and independent
of kernel conditioning.} \textbf{(a)} Designed-task accuracy advantage
($\Nph = 3$, $\Mmod = 12$, $n = 180$) under multinomial resampling of
every outcome distribution at $S$ shots, with task labels reconstructed
from the noisy kernels at each budget. The advantage at
$S = 3 \times 10^{4}$ (the per-image sample budget of
ref.~\cite{rambach2026}) matches the exact-arithmetic value.
\textbf{(b)} Geometric difference and designed-task advantage at matched
$n = 180$ for the rank-limited ($\Mmod = 12$, $\Dout = 220$) and
unconstrained ($\Mmod = 20$, $\Dout = 1140$) configurations: $\ggeo$
decreases by a factor of 5.6 while the task-level advantage is unchanged
within error.}
\label{fig:robustness}
\end{figure}

\subsection{Sampling Budgets}
Feature estimation in hardware is limited by photon-counting statistics.
We therefore repeat the full pipeline under multinomial resampling of
every outcome distribution, reconstructing the kernels, the adversarial
task labels, and the readout from the sampled data alone at each budget
$S$. The designed-task advantage is stable from $S = 10^{3}$ to
$10^{6}$ shots, and at $S = 3\times10^{4}$ (the per-image budget of
the experiments of ref.~\cite{rambach2026}) it is $+0.244 \pm 0.112$,
indistinguishable from the exact value (measured at the $\Mmod=12$
configuration; Fig.~\ref{fig:robustness}b shows the task-level advantage
is configuration-insensitive).
The geometric separation is thus accessible at current experimental
sample budgets, not only in exact arithmetic.

\subsection{Conditioning and the Meaning of $\ggeo$}
The converged $\ggeo$ is a conditioning-sensitive quantity: across sample
sizes $n = 60$--$180$ the product $\ggeo \cdot \lambda_{\min}(\Kc)$ is
approximately constant ($\approx 4 \times 10^{-3}$, spanning a 27-fold
range of $\ggeo$), because the unregularized geometric difference is
attained in the tail directions of the empirical classical kernel. Its
absolute value therefore depends on how deeply the finite-sample Gram
matrix resolves that tail, and all comparisons in this work are made at
matched $n$. The task-level advantage is not subject to this ambiguity:
between the rank-limited ($\Mmod = 12$, $\Dout = 220$) and unconstrained
($\Mmod = 20$, $\Dout = 1140$) configurations at matched $n = 180$,
$\ggeo$ decreases by a factor of 5.6 while the designed-task advantage
is unchanged within error
($+0.248 \pm 0.075$ vs.\ $+0.207 \pm 0.044$;
Fig.~\ref{fig:robustness}b). The learnable effect and the conditioning
artifact are dissociated by direct measurement.

Uniform loss does not affect any of these results: a uniform per-mode
efficiency $\eta$ rescales every collision-free $\Nph$-photon outcome by
$\eta^{\Nph}$ and cancels exactly in the renormalized post-selected
distributions. Mode-dependent loss is absorbed, at the level of
statistics, into an effective interferometer. Multiphoton emission noise ($g^{(2)}$) is not modelled; because it perturbs the input state itself rather than the channel, it is the imperfection most likely to modify the geometric quantities at fixed $V$, and its characterization is a priority for hardware follow-up.

% ================================================================
\section{Learnability Is Gated by Kernel--Task Alignment}
\label{sec:gating}

Throughout this section we report the unconstrained configuration
($\Mmod=20$, $\Dout=1140$); the rank-limited configuration gives
statistically indistinguishable gaps (§V.B), and out-of-sample (OOS)
labeling is used unless stated.

\begin{figure}[htbp]
\centering
\includegraphics[width=\linewidth]{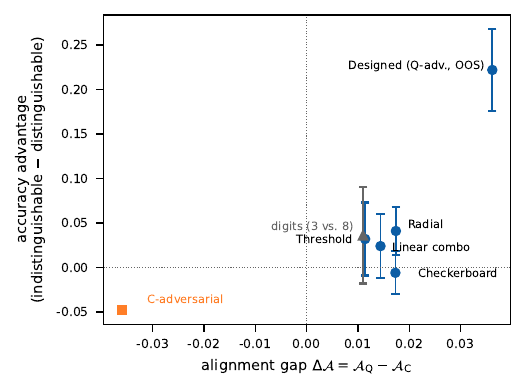}
\caption{\textbf{The learnable advantage is gated by kernel--task
alignment.} Accuracy advantage of the indistinguishable- over the
distinguishable-photon kernel against the alignment gap
$\Delta\alignK = \alignK_{\mathrm{Q}} - \alignK_{\mathrm{C}}$, for
adversarially designed and natural tasks
($\Mmod = 20$, $\Nph = 3$, $\Dout = 1140$, $n = 180$). The advantage
separates at the extremes of alignment; the fine-grained ordering among
natural tasks is not resolved at these sample sizes (Supplementary
Note~8).}
\label{fig:doseresponse}
\end{figure}

\begin{table}[t]
\centering
\caption{\textbf{Learnable advantage across task settings.} Accuracy
advantage of the indistinguishable- over the distinguishable-photon
kernel, with kernel--target alignments
$\alignK_{\mathrm{Q}}, \alignK_{\mathrm{C}}$ and their gap
$\Delta\alignK$, at $\Mmod = 20$, $\Nph = 3$, $\Dout = 1140$,
$n = 180$ (mean $\pm$ s.d.\ over 10 interferometer realizations and 10
data splits). Natural-task $p$-values from two-sided paired tests.}
\label{tab:doseresponse}
\renewcommand{\arraystretch}{1.2}
\begin{tabular}{@{}lcccc@{}}
\toprule
Setting & $\alignK_{\mathrm{Q}}$ & $\Delta\alignK$ &
Accuracy gap & $p$ \\
\midrule
Designed (Q-adv., OOS) & 0.067 & $+$0.036 &
$+0.222 \pm 0.046$ & $<10^{-4}$ \\
Designed (C-adversarial) & --- & (neg.) &
$-0.048$ & --- \\
Radial             & 0.094 & $+$0.017 & $+0.041 \pm 0.027$ & 0.001 \\
Threshold          & 0.104 & $+$0.011 & $+0.032 \pm 0.041$ & 0.041 \\
Linear combination & 0.078 & $+$0.014 & $+0.024 \pm 0.036$ & 0.082 \\
Checkerboard       & 0.049 & $+$0.017 & $-0.006 \pm 0.024$ & 0.501 \\
Digits (3 vs.\ 8)  & ---   & ---      & $+0.036 \pm 0.054$ & --- \\
\bottomrule
\end{tabular}
\end{table}

Geometry bounds what is learnable; alignment decides what is learned.
We measure the alignment of each kernel with a labeling $y \in
\{\pm 1\}^{n}$ by the standard kernel--target
alignment~\cite{cristianini2001},
\begin{equation}
\alignK(K, y) \;=\;
\frac{\langle K,\, y y^{\top} \rangle_{\mathrm{F}}}
{\lVert K \rVert_{\mathrm{F}}\, \lVert y \rVert^{2}},
\label{eq:alignment}
\end{equation}
and construct tasks that maximize the geometric difference by the
protocol of ref.~\cite{yin2025}: the labeling is the leading generalized
eigenvector of $\Kq^{1/2}(\Kc + \varepsilon\mathbbm{1})^{-1}\Kq^{1/2}$,
binarized at its median. To exclude any circularity from the
construction, the out-of-sample (OOS) variant derives the separating
direction from the training split alone and labels test points through
the kernel expansion; the reverse construction builds tasks adversarial
to $\Kc$ instead. Natural tasks are label families defined on the data
manifold of the inputs (Methods and Supplementary Note~3).

The designed-task advantage is $+0.222 \pm 0.046$ with OOS labels, and
the reverse construction favors the distinguishable-photon kernel
($-0.048$): the machinery carries no bias toward either map. Against
this, the natural-task advantages are small and graded
(Table~\ref{tab:doseresponse}): from $+0.041 \pm 0.027$ on the radial
family down to $-0.006 \pm 0.024$ on the checkerboard family, with the
real digits task at $+0.036 \pm 0.054$. Pooling over interferometer
realizations, the designed--natural separation is $+0.187$ (paired
$t = 8.69$, $p < 10^{-4}$).

Across the settings examined, the advantage separates at the extremes:
the designed task, with the largest alignment gap, realizes the full
advantage, while tasks with zero or negligible alignment realize none;
the random-label control produces $+0.009$, indistinguishable from zero.
The fine-grained ordering among natural tasks, however, is not resolved
at these sample sizes: in a pre-registered test across thirteen further
task families (10 realizations $\times$ 10 splits), the alignment gap
predicted the sign of the advantage in 9 of 13 cases (registered
criterion: at least 10 of 14), and the rank correlation between
alignment gap and advantage was not significant ($\rho = 0.027$,
$p = 0.93$). The natural-task alignment gaps ($\Delta\alignK =
0.003$--$0.037$) lie within the realization noise of the accuracy
differences ($\pm 0.02$--$0.045$). The gating claim, that geometry
without alignment yields no advantage, is supported at the extremes and
by both controls; the graded ordering within the natural regime is a
resolution limit, not an established structure (Supplementary Note~8).
The weighted reading $\alignK_{\mathrm{Q}} \cdot \Delta\alignK$ orders
the six original settings consistently but is unsupported by the
expanded battery and is retained only as the observation that motivated
it (Supplementary Note~5).

The advantage does not grow with the size of the quantum system. At
matched $\Mmod = 20$ and $n$, the designed-task advantage is
$+0.305 \pm 0.152$, $+0.207 \pm 0.044$, and $+0.165 \pm 0.059$ for
$\Nph = 2, 3, 4$, constant within error bars across a 25-fold range of
feature dimension, consistent with alignment, not Hilbert-space size,
governing learnability.

Finally, the separation is specific to the photonic maps rather than to
high-dimensional random features in general. Random Fourier features~\cite{rahimi2007,huang2006,saade2016,pierangeli2021} matched in dimension and preprocessing do not
recover the designed-task advantage under adversarial construction
against either map. On tasks designed against the classical kernel,
the quantum kernel attains 0.761 accuracy where RFF attains 0.583.
On tasks designed against the RFF map, both photonic kernels exceed RFF
(0.883 for quantum and 0.878 for classical, versus 0.439 for RFF),
while the quantum and classical kernels remain indistinguishable from
each other. Both photonic statistics share structure that generic random
features lack; the interference-specific contribution is the quantum
versus classical photonic separation.

% ================================================================
\section{Reconciling the Experimental Record}
\label{sec:unification}

\begin{table*}[t]
\centering
\caption{\textbf{The mixed experimental record through the lens of
kernel geometry and task alignment.}}
\label{tab:unification}
\renewcommand{\arraystretch}{1.3}
\begin{tabular}{@{}p{2.4cm}p{3.6cm}p{4.4cm}p{4.4cm}@{}}
\toprule
Experiment & Setting & Reported result & Account given here \\
\midrule
Joly et al.~\cite{joly2025} &
Multimode fibre, $\Nph = 2$, 22 detectors, MNIST 0--1, single
transmission matrix &
No resolved distinguishability effect; higher feature rank for
indistinguishable photons &
Easiest binary task (no alignment); rank-limited features; single
interferometer draw. Geometry present, unexploited (held-out
prediction, Supplementary Note~6) \\[2pt]
Di Bartolo et al.~\cite{dibartolo2026} &
4-mode PIC, $\Nph = 2$, feedback, temporal tasks &
Advantage on parity, NARMA, Mackey--Glass; none on linear memory &
Task nonlinearity spans the alignment axis; the linear-recall null is
the control. Expressivity axis is interference; memory axis is feedback \\[2pt]
Rambach et al.~\cite{rambach2026} &
$\Nph = 3$, $\Mmod = 12$--$24$, quantum processor, MNIST/MedMNIST &
Indistinguishability aids training accuracy; test accuracy nearly
unchanged &
Expressivity (geometry) improves trainability; generalization gated by
the weak alignment of their natural tasks \\[2pt]
Yin et al.~\cite{yin2025} &
$\Nph = 2$, 6 modes, overlap kernels, designed tasks &
Quantum kernel outperforms classical on designed tasks &
Designed alignment realizes the separation; the overlap setting is
well-conditioned \\
\bottomrule
\end{tabular}
\end{table*}

The four experiments of Table~\ref{tab:unification} were performed on
different platforms, photon numbers, and tasks; we suggest they are
better viewed as sampling different regions of a single landscape. Joly
et al.\ and Di Bartolo et al.\ form a natural experiment at $\Nph = 2$:
the same physical resource, opposite outcomes, with the differentiator
being task structure, an easy static classification in one case,
strongly nonlinear temporal tasks in the other. Within the Di Bartolo
experiment itself, the advantage is graded by task demand, and vanishes
on the one task (linear memory recall) for which alignment with the
interference-specific directions should be negligible. Rambach et al.\
resolve the same distinction as a train--test gap: the geometric
resource improves trainability on all tasks, while generalization
follows alignment.

As a prospective-style test of this account, we computed the
framework's predictions for the configuration of ref.~\cite{joly2025}
(their photon number, detector count, task, and sample size) without
reference to their results. The framework predicts a geometric
difference of $\ggeo = 4.4 \pm 0.5$ at full indistinguishability but an
alignment gap of $\Delta\alignK = +0.008$ for their task, and hence no
resolvable distinguishability advantage at their sample size: the
predicted effect of indistinguishability ($+0.015$) is smaller than the
predicted interferometer-to-interferometer spread ($\pm 0.017$). This
reproduces the structure of their reported null: the published
effect ($+0.014$) is likewise smaller than their variability
($\pm 0.042$), and the predicted effect magnitude agrees with the
published value to within 10\%. The absolute accuracies differ by a
constant $+0.04$ across the dial, attributable to readout and
detector-noise differences; the $\alpha$-dependence, which is the
object of the prediction, is unaffected (Supplementary Note~6).

Our account is complementary to the Fisher-information interpretation
offered in ref.~\cite{dibartolo2026}, where indistinguishable states
increase the sensitivity of output probabilities to input variation.
Sensitivity to input variation, in kernel terms, is variance spread
across directions of the outcome space; the Fisher-information account describes the same variance structure in parameter space; the present account describes it in kernel space.

% ================================================================
\section{Discussion}

\begin{figure*}[htbp]
\centering
\begin{minipage}{0.96\linewidth}
\hrule\vspace{0.8em}
\textbf{Box 1: Pre-experimental protocol for interference in
photonic reservoir computing}

\vspace{0.5em}
Given a candidate task family and system parameters
 $(\Mmod, \Nph, d)$:
\begin{enumerate}
\setlength{\itemsep}{2pt}
\item \textbf{Estimate the kernels.} On a small pilot set of
unlabeled inputs $\{x_i\}_{i=1}^{n}$ (we find $n \sim 100$ sufficient), compute the outcome distributions for indistinguishable
and distinguishable photons, Eqs.~(2)--(3), and form the Gram
matrices $\Kq, \Kc$, Eq.~(5). In hardware, this requires
 $\mathcal{O}(n)$ circuit evaluations at a modest shot budget
($S \approx 3\times10^{4}$ suffices; Fig.~3a).
\item \textbf{Compute the geometric difference.}
 $\ggeo = \sqrt{\lambda_{\max}(\Kq^{1/2}(\Kc +
\varepsilon\mathbbm{1})^{-1}\Kq^{1/2})}$, with $\varepsilon \ll
\lambda_{\min}(\Kc)$. Compare only $\ggeo$ values computed at
matched $n$ (Supplementary Note 1).
\item \textbf{Compute task alignment.} For each candidate labeling
family $y$, evaluate
 $\alignK_{\mathrm{Q}} = \alignK(\Kq, y)$ and
 $\alignK_{\mathrm{C}} = \alignK(\Kc, y)$, Eq.~(8). Label families
are cheap to enumerate; the kernels are computed once.
\item \textbf{Decision.} Interference is expected to help only if $\ggeo$ is large and $\alignK_{\mathrm{Q}}$ exceeds $\alignK_{\mathrm{C}}$ on the task family. Otherwise, high-visibility sources offer no expected benefit for this task.
\end{enumerate}
\vspace{0.8em}\hrule
\end{minipage}
\captionsetup{type=figure}
\vspace{-0.5em}
\end{figure*}

The results assemble into a pre-experimental protocol. Before committing
quantum hardware, one computes the two Gram matrices (exactly in
simulation, or from a modest shot budget in hardware~\cite{xiong2025}) and evaluates the
geometric difference $\ggeo$ and the alignments
$\alignK_{\mathrm{Q}}, \alignK_{\mathrm{C}}$ against the task family of
interest. Interference is worth its experimental cost only when both are
favorable. The geometric difference certifies that the quantum kernel contains directions absent from the classical map. Whether a given task makes use of them is a separate question, settled by alignment. In the record examined here, the experiments that found a
task-level advantage are those whose tasks aligned; those that did not
are those whose tasks did not. Candidate label families can be evaluated
cheaply; the protocol ranks them before any labeling cost is incurred. We note that the reconciliation offered here is retrospective: the framework was developed with knowledge of the four experimental outcomes it accounts for. Its prospective test, computing $\ggeo$ and alignment for a task family and visibility in advance of measurement, is the definitive validation, and we identify it as the immediate follow-up. As a first step in this
direction, the framework reproduces the reported null of
ref.~\cite{joly2025} when applied blind to their configuration
(Supplementary Note~6), and the pre-registered task battery of
Supplementary Note~8 was likewise specified before its results existed.

The same configuration admits a sharper analysis of the interior optimum
reported in ref.~\cite{joly2025}. We prove that the model complexity
$s_{K(\vis)}(y)$ is convex in $\vis$ for every labeling: by
Cauchy--Schwarz on the two feature embeddings, the cross-statistics term
obeys $s_{\mathrm{X}} \le \sqrt{s_{\mathrm{Q}} s_{\mathrm{C}}}$, so the
quadratic coefficient of $s_{K(\vis)}$ satisfies
$s_{\mathrm{Q}} + s_{\mathrm{C}} - 2s_{\mathrm{X}} \ge
(\sqrt{s_{\mathrm{Q}}} - \sqrt{s_{\mathrm{C}}})^2 \ge 0$
(Supplementary Note~7, Proposition~2). Interior optima are therefore
impossible at the level of kernel geometry for any task. The measured
cross-alignment at their configuration ($\alignK_{\mathrm{X}} = 0.071$,
below both $\alignK_{\mathrm{Q}} = 0.080$ and
$\alignK_{\mathrm{C}} = 0.072$) confirms the mechanism is absent there.
Finally, a power analysis of their published curve shows the reported
edge (a peak exceeding the fully indistinguishable endpoint by 0.002
against a transmission-matrix spread of $\pm 0.042$) would require
approximately 800 matrix draws to resolve at $1\sigma$; their
experiment used 100, sufficient to establish the monotone rise but
roughly twenty-fold too noisy to distinguish an interior optimum from
an endpoint maximum. The reported optimum is thus either a
readout-level effect or within sampling variability; it cannot be
geometric.

Partial distinguishability, usually treated as an imperfection, appears
here as a continuous resource. The retention at hardware-typical
visibility is high, and optimal task performance need not occur at full
indistinguishability. Interior optima in distinguishability are provably
non-geometric (Section~\ref{sec:unification}); whether readout-level
mechanisms can produce them, as suggested by the partial-coherence
results of ref.~\cite{dong2024}, is an open question.

More broadly, the geometric quantities studied here connect the photonic reservoir literature~\cite{xia2024,nakajima2021} to the kernel-theoretic understanding of quantum
machine learning~\cite{innocenti2023,biamonte2017,schuld2021,huang2021,liu2021,kubler2021,cerezo2021,mcclean2018,jerbi2023,caro2022,wang2021}: the
concentration warnings raised for large quantum
kernels~\cite{thanasilp2024} appear here as the conditioning sensitivity
of $\ggeo$, and the distinction between expressivity and generalization
that recurs across experiments is the distinction between geometry and
alignment.

\paragraph*{Limitations} Our results are numerical, at system sizes
that remain classically simulable; they bracket the experimental regime
through shot-noise and visibility modeling but do not replace hardware
validation. The adversarial tasks are synthetic constructions built by
the protocol of ref.~\cite{yin2025}, answered here by out-of-sample
labeling, the reverse construction, and the natural-task arm, but
synthetic nonetheless. We do not model temporal feedback; within the
experiments we account for, the distinguishability effect is confined to
the expressivity axis, which our framework addresses directly, while the
memory axis is feedback-governed~\cite{kobayashi2024} and irrelevant to interference effects in the data
of ref.~\cite{dibartolo2026}. The natural-task families are structured
label functions on a single data manifold; the fine-grained alignment
ordering among natural tasks is not resolved at current sample sizes
(Supplementary Note~8), and the weighted-alignment reading is post-hoc
and unsupported by the expanded battery. The absolute value of $\ggeo$ is
conditioning-sensitive and sample-size dependent; all comparisons are
made at matched configuration.

The pre-experimental protocol is computationally bound: estimating the quantum kernel $\Kq$ requires evaluating matrix permanents, a \#P-hard task at scale~\cite{aaronson2011,gurvits2011,clifford2018}. The protocol therefore operates in the engineering regime (small-to-midscale systems, such as the $\Nph=3$ simulations performed prior to hardware runs in ref.~\cite{rambach2026}), where it determines whether allocating quantum hardware time is justified~\cite{preskill2018}. For large-scale systems ($\Nph \ge 50$)~\cite{zhong2020,madsen2022}, the protocol cannot be evaluated classically; it guides the decision up to the classically simulable frontier, and beyond this limit, predicting utility relies on extrapolating alignment heuristics, which remains an open problem.

% ================================================================
\section*{Methods}
\label{sec:methods}

\subsection*{Feature Maps and Kernels}
Simulations use the encoding of Eq.~\eqref{eq:encoding} with independent
Haar-random $U_{\mathrm{pre}}$ and $U_{\mathrm{res}}$. Unless stated
otherwise, the flagship geometry configuration is $\Mmod = 12$,
$\Nph = 3$, $d = 12$, $n = 100$ ($\Dout = 220$), and the flagship
accuracy configuration is $\Mmod = 20$, $\Nph = 3$, $d = 12$, $n = 180$
($\Dout = 1140$). Outcome probabilities are computed exactly from
Eq.~\eqref{eq:partial} with Ryser's algorithm for the permanent.
Features are the collision-free outcome distributions, post-selected on
$\Nph$ detection events and renormalized; the geometric analysis uses
the raw kernels of Eq.~\eqref{eq:kernels}, and the operational
standardized kernels as a cross-check (Supplementary Note~1). The
visibility convention is $\vis = I^{2}$ throughout.

\subsection*{Geometric Difference and Alignment}
The geometric difference follows Eq.~\eqref{eq:gdef} in the square-root
convention of ref.~\cite{yin2025}, with the regularization $\varepsilon$
taken at least three orders of magnitude below the smallest eigenvalue
of $\Kc$; convergence curves and the conditioning-pinning relation
$\ggeo \cdot \lambda_{\min}(\Kc) \approx \mathrm{const.}$ are documented
in Supplementary Note~1. Retention is defined by Eq.~\eqref{eq:retention}
and computed per interferometer realization. Kernel--target alignment
follows Eq.~\eqref{eq:alignment}~\cite{cristianini2001}.

\subsection*{Task Construction}
Adversarial tasks follow ref.~\cite{yin2025}: the labeling is the leading
generalized eigenvector of
$\Kq^{1/2}(\Kc + \varepsilon\mathbbm{1})^{-1}\Kq^{1/2}$ mapped through
$\Kq^{1/2}$ and binarized at the median. In the out-of-sample variant,
the direction is derived from the training split alone and test labels
are assigned by the kernel expansion. The reverse construction exchanges
the roles of $\Kq$ and $\Kc$. Natural tasks are label families
(threshold, checkerboard, radial, linear combination) defined on the
input manifold, together with the 3-vs-8 digits classification; exact
generators are given in Supplementary Note~3.

\subsection*{Evaluation Protocol}
Readouts are kernel ridge classifiers with the regularization selected
by cross-validation on the training split. All results average over 10
Haar interferometer realizations and 10 stratified data splits; the
gating separation is assessed by a paired $t$-test on per-realization
differences between designed and natural gaps. Shot-noise experiments
multinomially resample every outcome distribution at budget $S$ and
reconstruct kernels, labels, and readout from the sampled data alone;
outcomes with probability below $1/(2S)$ are trimmed. Random Fourier feature baselines~\cite{rahimi2007,huang2006,saade2016,pierangeli2021} are matched to $\Dout$ features and
processed through the identical pipeline.

\subsection*{Numerical Integrity}
Permanents are validated against brute-force enumeration; both outcome
statistics sum to unit norm including collision terms; the
distinguishable and indistinguishable statistics coincide at $\Nph = 1$;
the partial-distinguishability construction of Eq.~\eqref{eq:partial}
reproduces both endpoints to machine precision; and the full machinery
reproduces the overlap-kernel setting of ref.~\cite{yin2025}
(Supplementary Note~4). All
results are regenerated from a single version-controlled pipeline.

% ================================================================
\section*{Data and Code Availability}
The simulation code, frozen pipeline, and all numerical data supporting
the findings are available at
\url{https://anonymous.4open.science/r/photonic-reservoir-kernel-geometry-3739}.
The de-anonymized repository and commit hash will be provided upon
publication.

\section*{Acknowledgements}
The authors acknowledge the facilities of the Indian Institute of
Information Technology Dharwad. No external funding was received for
this work. The authors used open-weight large-language-model tools for
minor assistance with code development and English grammar checking. All research decisions, results,
and final text are the authors' own.

\section*{Author Contributions}
O.K.: Conceptualization, Methodology, Software, Investigation, Data
curation, Visualization, Writing -- original draft. A.B.H.:
Supervision, Writing -- review \& editing.

\section*{Competing Interests}
The authors declare no competing interests.

%\bibliographystyle{IEEEtranN}
%\bibliography{projectx}

% ================================================================
% REFERENCES -- converted from projectx(1).bib for direct inclusion
% No separate .bib file or BibTeX run is required.
% ================================================================

\clearpage
\twocolumn[{
\begin{center}
{\LARGE\bfseries Supplementary Information\par}
\vspace{6pt}
{\large for ``When Does Interference Help Learning? Kernel Geometry as a Pre-Experimental Test for Photonic Reservoir Computing''\par}
\vspace{10pt}
{\normalsize Oishik Kar$^{1,\ast}$, Aswath Babu H$^{2}$\par}
{\small $^{1}$Dept.\ of CSE, $^{2}$Dept.\ of Arts, Science and Design, IIIT Dharwad, India\par}
\vspace{1.2em}
\end{center}
}]

\section*{Note 1: Convergence and conditioning of the geometric
difference}

\subsubsection*{Regularization}
\begin{figure}[H]
\centering
\includegraphics[width=\linewidth]{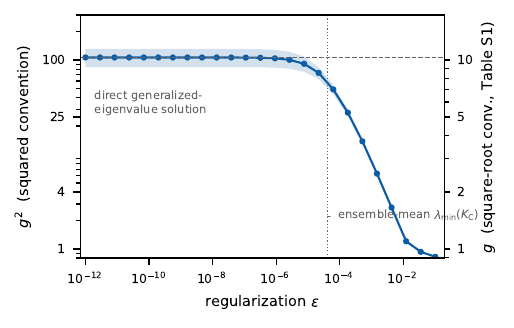}
\caption{\textbf{Convergence of the geometric difference with
regularization.} $\ggeo$ versus $\varepsilon$ at the flagship
geometry configuration (mean over ten interferometer realizations;
shaded band, $\pm$ one standard deviation; squared convention on
the left axis, with the square-root values of
Table~\ref{tab:eps} on the right axis). The vertical line marks
the ensemble-mean $\lambda_{\min}(\Kc) \approx 4\times10^{-5}$: for
 $\varepsilon \ll \lambda_{\min}(\Kc)$ the regularization is inert
and $\ggeo$ equals the direct generalized-eigenvalue solution
(dashed horizontal line).}
\label{fig:eps}
\end{figure}

The geometric difference is computed as
 $\ggeo(\Kq,\Kc) = \sqrt{\lambda_{\max}(\Kq^{1/2}
(\Kc + \varepsilon\identity)^{-1} \Kq^{1/2})}$ with numerical
regularization $\varepsilon$. Because the unregularized quantity is
attained in the tail directions of $\Kc$, its value depends on how
deeply $\varepsilon$ probes that tail; we therefore report
convergence explicitly. Table~\ref{tab:eps} shows $\ggeo$ versus
 $\varepsilon$ at the flagship geometry configuration
($\Mmod{=}12$, $\Nph{=}3$, $d{=}12$, $n{=}100$; mean over ten
interferometer realizations, square-root convention). The value is
flat below $\varepsilon = 10^{-6}$ and matches the direct
generalized-eigenvalue solution, consistent with
 $\lambda_{\min}(\Kc) \approx 4\times10^{-5}$: once
 $\varepsilon \ll \lambda_{\min}(\Kc)$, the regularization is inert.
No eigenvalue of $\Kc$ approaches machine precision.
Figure~\ref{fig:eps} displays the convergence curve.

\begin{table}[H]
\centering
\caption{Convergence of $\ggeo$ with regularization. Flagship
configuration; mean over ten realizations.}
\label{tab:eps}
\setlength{\tabcolsep}{4pt}
\begin{tabular}{lcccccc}
\toprule
 $\varepsilon$ & $10^{-12}$ & $10^{-10}$ & $10^{-8}$ & $10^{-6}$ &
 $10^{-4}$ & $10^{-2}$ \\
\midrule
 $\ggeo$ & 10.32 & 10.32 & 10.32 & 10.22 & 6.99 & 1.10 \\
\bottomrule
\end{tabular}
\end{table}

\subsubsection*{Conditioning-pinning}
Across $n = 60$--$180$ at fixed configuration,
 $\ggeo \cdot \lambda_{\min}(\Kc)$ is approximately constant
($\approx 4\times10^{-3}$ in the squared convention, spanning a
27-fold range of $\ggeo$). This follows from the generalized
eigenvalue characterization: the maximal eigenvalue of
 $\Kq^{1/2} \Kc^{-1} \Kq^{1/2}$ is attained on the eigenvector of $\Kc$ with the smallest eigenvalue, so
 $\ggeo^{2} \sim \langle v_{\min}, \Kq v_{\min}\rangle /
\lambda_{\min}(\Kc)$ with an approximately $n$-independent numerator.
The absolute value of $\ggeo$ is therefore a conditioning readout;
all main-text comparisons are made at matched $n$ and configuration,
and the task-level results are shown there to be insensitive to it by
direct measurement (dissociation experiment, main-text Fig.~3b).
Table~\ref{tab:gn} documents the sample-size dependence.

\begin{table}[H]
\centering
\caption{Sample-size dependence. Flagship configuration,
representative realization, squared convention.}
\label{tab:gn}
\setlength{\tabcolsep}{3.5pt}
\footnotesize
\begin{tabular}{lccccc}
\toprule
 $n$ & 60 & 80 & 100 & 150 & 180 \\
\midrule
 $\ggeo^{2}$ & 39.1 & 65.5 & 101.4 & 411.5 & 1052.2 \\
 $\lambda_{\min}(\Kc)$ & $1.0\!\times\!10^{-4}$ &
 $6.7\!\times\!10^{-5}$ & $3.7\!\times\!10^{-5}$ &
 $1.1\!\times\!10^{-5}$ & $4.0\!\times\!10^{-6}$ \\
 $\ggeo^{2}\lambda_{\min}$ & $3.9\!\times\!10^{-3}$ &
 $4.4\!\times\!10^{-3}$ & $3.8\!\times\!10^{-3}$ &
 $4.5\!\times\!10^{-3}$ & $4.2\!\times\!10^{-3}$ \\
\bottomrule
\end{tabular}
\end{table}

\subsubsection*{Kernel-convention cross-check}
The main text uses the raw kernels of Eq.~(5). The standardized
feature kernels of the operational pipelines of
refs.~\cite{sakurai2025,rambach2026} show the same spectral contrast:
effective dimension at 90\% cumulative variance of 61
(indistinguishable) vs.\ 33 (distinguishable), leading-eigenvalue
ratio in the same direction. The $z$-scored $\ggeo$ at
 $\varepsilon = 10^{-3}$ is suppressed relative to the raw value, as
expected from variance normalization; both conventions are monotone
in $\vis$ with identical retention orderings.

% =====================================================================
\section*{Note 2: Exponent analysis of the dial}

\subsubsection*{Small visibility}
At small $\vis$, $\ggeo(\vis) - 1$ is linear in $\vis$ for every
photon number examined; effective exponents
 $k_{\vis} = \mathrm{d}\log(\ggeo-1)/\mathrm{d}\log\vis$ over the
fitted window $\vis \in [0.01, 0.16]$ are given in
Table~\ref{tab:kv}. The leading behavior admits a perturbative
account: in the distinguishability expansion of the outcome
probabilities, the correction to the fully distinguishable statistics
is a sum over permutations grouped by cycle structure; transpositions
enter with weight $I^{2} = \vis$ and dominate at leading order for
every $\Nph$, longer cycles being suppressed by higher powers. For
 $\Nph{=}2$ the expansion is exact: two-photon outcome probabilities
are affine in $\vis$, so $\ggeo - 1$ is linear in $\vis$ plus
quadratic corrections, and the fitted exponent of 1.057 reflects
window bias rather than deviation from linearity. The approach of
 $k_{\vis}$ to 1 from above across $\Nph = 2, 3, 4$ is consistent with
transposition dominance universal in $\Nph$.

\begin{table}[H]
\centering
\caption{Small-visibility exponents. $\Nph{=}2$: $\Mmod{=}20$,
 $n{=}80$; $\Nph{=}3$: $\Mmod{=}12$, $n{=}100$; $\Nph{=}4$:
 $\Mmod{=}20$, $n{=}60$ (small-$\vis$ window only).}
\label{tab:kv}
\setlength{\tabcolsep}{5pt}
\begin{tabular}{lccc}
\toprule
 $\Nph$ & 2 & 3 & 4 \\
\midrule
 $k_{\vis}$ & 1.057 & 1.088 & 1.117 \\
 $R^{2}$ & 1.000 & 0.999 & 0.998 \\
\bottomrule
\end{tabular}
\end{table}

\subsubsection*{Full dial}
Over $\vis \in [0.1, 1]$, the effective exponent grows with photon
number: $k_{\vis} = 1.16$ at $\Nph{=}2$ and $1.22$ at $\Nph{=}3$ ($\Nph{=}4$ was computed over the small-$\vis$ window only, at
 $\Mmod{=}20$, $\Dout{=}4845$), reflecting the accumulation of
higher-cycle interference terms. The window-fitted exponent (1.16 at
$\Nph{=}2$) and the analytic slope of Proposition~1 in Note~7 (1.67 at
$\Nph{=}2$) differ because the window fit averages the convex bend
across the full fitted range; both quantities are correct and measure
different things. The $\Nph{=}4$ dial was recomputed on the frozen
recipe (3 seeds; $\ggeo(1) = 4.05 \pm 0.16$), replacing an earlier run
that could not be reproduced from a saved artifact. Two caveats: the full-range
exponents are effective (window-dependent) quantities, and the
 $\Nph{=}2$ and $\Nph{=}3$ curves were computed at different
 $(\Mmod, n)$, so the comparison is indicative rather than
matched-configuration. The small-$\vis$ linearity is the robust
statement. Endpoint validation: the partial-distinguishability
construction reproduces $\Kq$ at $\vis{=}1$ and $\Kc$ at $\vis{=}0$ to
machine precision (entrywise deviation below $10^{-15}$) at every
configuration used (Note 4).

% =====================================================================
\section*{Note 3: Natural task families and generators}

All natural tasks are label functions on the input manifold
 $[0,1]^{d}$, evaluated on the same uniform input sets used for the
kernels, except the digits task. Labels are binarized at the family
median where required. Table~\ref{tab:families} gives the generators
exactly as implemented in the frozen pipeline.

\begin{table}[H]
\centering
\caption{Natural task families. $x = (x_0, \dots, x_{d-1}) \in
[0,1]^{d}$, $d = 12$.}
\label{tab:families}
\setlength{\tabcolsep}{4pt}
\footnotesize
\begin{tabular}{@{}p{1.9cm}p{5.4cm}@{}}
\toprule
Family & Labeling $y(x)$ \\
\midrule
Threshold & $\operatorname{sign}(x_0 - 0.5)$ \\
Checkerboard & $\operatorname{sign}\!\big(\sin(4\pi x_0)\,
\sin(4\pi x_1)\big)$ \\
Radial & $\operatorname{sign}\!\big(\sum_{i=0}^{2}
(x_i - 0.5)^{2} - 0.25\big)$ \\
Linear comb. & $\operatorname{sign}\!\big(0.4\,x_0 + 0.3\,x_1 +
0.2\,x_2 + 0.1\,x_3 - 0.5\big)$ \\
Digits (3v8) & UCI digits class labels, PCA to $d$ components,
scaled to $[0,1]^{d}$ \\
\bottomrule
\end{tabular}
\end{table}

The radial family is a sphere of radius $0.5$ centered at the
manifold center in the first three coordinates; the linear family
uses fixed weights $(0.4, 0.3, 0.2, 0.1)$ over the first four
coordinates with threshold $0.5$; the checkerboard frequency
corresponds to two oscillations per unit interval in each of the
first two coordinates. The families span task structure on the
manifold: a single-coordinate split, a high-frequency product
structure, a smooth geometric boundary, and a generic linear
boundary. The checkerboard family realizes the lowest absolute
alignments of all settings ($\alignK_{\mathrm{Q}} = 0.049$),
consistent with high-frequency structure being weakly represented in
the photonic feature maps.

\subsubsection*{Adversarial tasks}
Following ref.~\cite{yin2025}: the labeling is the leading generalized
eigenvector of $\Kq^{1/2}(\Kc + \varepsilon\identity)^{-1}
\Kq^{1/2}$, mapped through $\Kq^{1/2}$, binarized at the median. In
the out-of-sample (OOS) variant, the eigenvector is computed from the
training split alone; test labels are assigned by the sign of the
kernel expansion $\sum_{i \in \mathrm{train}} \alpha_i
\Kq(x, x_i)$ restricted to training points. The reverse construction
exchanges $\Kq$ and $\Kc$.

\subsubsection*{Shot-noise protocol}
Sampling-limited experiments are simulated by multinomial
resampling of every collision-free outcome distribution at budget
 $S$: each input's exact distribution $P(\cdot \mid x)$ is replaced
by a draw $\hat{P} \sim \mathrm{Mult}(S, P(\cdot \mid x))$. The full
pipeline is then reconstructed from sampled data alone: (i) outcome
trimming: outcomes with exact probability below $1/(2S)$ are removed
\emph{before} renormalization of $\hat{P}$, since their sample
estimates carry no information at this budget; (ii) kernel
reconstruction: the Gram matrices are computed from the trimmed,
renormalized empirical distributions; (iii) label reconstruction:
adversarial task labels are re-derived from the noisy kernels, so
that the designed-task construction itself operates under the same
sampling budget as the readout (the out-of-sample protocol is
unchanged); (iv) readout: kernel ridge classification as below. The
standardization floor for the $z$-scored cross-check kernels is set
to $10^{-3}$ of the outcome-space standard deviation to prevent
amplification of undersampled outcomes. Each budget level is
evaluated over the full realization $\times$ split ensemble.

\subsubsection*{Random Fourier feature baseline}
The classical random-feature control follows
ref.~\cite{rahimi2007}: features
 $z_i(x) = \sqrt{2/D}\cos(w_i^{\top}x + b_i)$ with
 $w_i \sim \mathcal{N}(0, \gamma^{-2}\mathrm{I})$,
 $b_i \sim \mathrm{Uni}[0, 2\pi]$, implemented via \texttt{RBFSampler}
with the number of components matched to $\Dout$ and the same
random-state seeding as the interferometer realizations. The inputs
were standardized, and the bandwidth was fixed to $\gamma = 1.0$ (selected by cross-validation in preliminary runs). The baseline is
processed through the identical pipeline as the photonic maps:
identical trimming, standardization, readout, and split protocol.

\subsubsection*{Evaluation}
Kernel ridge classification, regularization by 3-fold
cross-validation on the training split; stratified 80/20 splits; 10
interferometer realizations $\times$ 10 data splits; gating assessed
by paired $t$-test on per-realization designed-minus-natural gap
differences.

% =====================================================================
\section*{Note 4: Numerical integrity}

All results are regenerated from a single version-controlled pipeline
state (repository and commit hash in the data availability statement
of the main text). The validation suite, executed before every
analysis run:

\begin{enumerate}
\setlength{\itemsep}{1pt}\setlength{\parskip}{0pt}
\item \textbf{Permanents:} Ryser implementation against brute-force
enumeration for complex $n = 2, 3, 4$, agreement to $10^{-9}$.
\item \textbf{Normalization:} both outcome statistics, including
collision terms with factorial denominators, sum to unit norm at
 $(\Nph,\Mmod) \in \{(2,4),(3,5)\}$.
\item \textbf{Physical limit:} for two indistinguishable photons
through a balanced beamsplitter the output is fully bunched
(Hong--Ou--Mandel) while distinguishable photons split binomially;
the implementation reproduces $(0.5, 0, 0.5)$ and $(0.25, 0.5,
0.25)$ to $10^{-12}$.
\item \textbf{Single-photon identity:} at $\Nph{=}1$ the two
statistics coincide exactly.
\item \textbf{Endpoint exactness:} the partial-distinguishability
construction reproduces $\Kq$ at $\vis{=}1$ and $\Kc$ at $\vis{=}0$ to
machine precision (below $10^{-15}$ entrywise) at every
configuration used.
\item \textbf{Calibration against experiment:} the full machinery,
run in the overlap-kernel setting of ref.~\cite{yin2025}, reproduces
the designed-task advantage of the indistinguishable kernel in that
setting (gap $+0.067$ in our implementation), confirming protocol
and convention alignment.
\end{enumerate}

\subsubsection*{Statistical protocol}
Uncertainties are standard deviations over the stated ensembles. The
gating separation is a paired test over per-realization differences;
an unpaired comparison would underestimate the separation because both
arms of each pair share the interferometer realization and data split.

% =====================================================================
\section*{Note 5: Weighted alignment reading}

The main text notes one instructive exception to the rank consistency
between the accuracy advantage and the alignment gap
 $\Delta\alignK$: the checkerboard family combines a sizeable
 $\Delta\alignK$ with no advantage, while carrying the lowest absolute
alignment of all settings. This motivates the weighted quantity
 $P = \alignK_{\mathrm{Q}} \cdot \Delta\alignK$, which orders all
settings consistently with the accuracy gap
(Table~\ref{tab:weighted}).

\begin{table}[H]
\centering
\caption{Weighted-alignment ordering. Same data as main-text Table 1.
The C-adversarial setting is consistent by construction.}
\label{tab:weighted}
\setlength{\tabcolsep}{4pt}
\footnotesize
\begin{tabular}{@{}lccc@{}}
\toprule
Setting & $\alignK_{\mathrm{Q}}$ & $P\,(\times10^{-3})$ & Gap \\
\midrule
Designed (OOS) & 0.0665 & 2.41 & $+0.222 \pm 0.046$ \\
Radial & 0.0940 & 1.64 & $+0.041 \pm 0.027$ \\
Threshold & 0.1039 & 1.18 & $+0.032 \pm 0.041$ \\
Linear comb. & 0.0782 & 1.13 & $+0.024 \pm 0.036$ \\
Checkerboard & 0.0490 & 0.85 & $-0.006 \pm 0.024$ \\
\bottomrule
\end{tabular}
\end{table}

\subsubsection*{Status}
This reading is post-hoc: identified after observing the checkerboard
exception, evaluated on the same settings it summarizes. We report it
as a hypothesis-generating observation, not an established criterion.
A proper validation would evaluate $P$ on held-out task families; we
leave this to future work. The main text draws no claim on $P$ beyond
its consistency. The expanded pre-registered battery (Note 8) does not
support the weighted reading's ordering claim; it is retained here only
as the observation that motivated the registration.

% =====================================================================
\section*{Note 6: Held-out prediction for the configuration of
\cite{joly2025}}

\subsubsection*{Protocol}
The framework was applied to the configuration of \cite{joly2025}
without reference to their reported results: two photons, 22
detection modes, their task (MNIST 0--vs--1, $8\times 8$), and
their sample size. The interferometer is modeled as an independent
complex-Gaussian random transmission matrix, the standard
idealization for their multimode fibre, in two layers
(64 phase-controlled SLM pixels $\to$ 290 guided fibre modes $\to$
22 detectors, with one independent matrix per photon); at
$\Nph = 2$ the $\alpha$-mixing of their Eq.~(1) coincides exactly
with the framework's visibility dial, $\alpha = \vis$.
Predictions are averaged over 100 transmission-matrix draws at
$10^{5}$ coincidences per input. The readout is kernel ridge
regression with cross-validated regularization on standardized
features.

\subsubsection*{Results}
Table~\ref{tab:heldout} compares the predicted and published
accuracies. The predicted effect of indistinguishability on
accuracy is $+0.015$ against a predicted TM-to-TM spread of
$\pm 0.017$; the published effect is $+0.014$ against variability
$\pm 0.042$. In both cases the effect lies below the realization
spread, reproducing the structure of the reported null. The
predicted geometric difference at full indistinguishability is
$\ggeo = 4.4 \pm 0.5$ ($n = 100$, raw kernels), and the predicted
alignment gap at their task is $\Delta\alignK = +0.008$
($\alignK_{\mathrm{Q}} = 0.334$ vs.\
$\alignK_{\mathrm{C}} = 0.326$), rising monotonically with
$\alpha$ as the accuracy does. The cross-kernel alignment at their
configuration is $\alignK_{\mathrm{X}} = 0.071$, below both
$\alignK_{\mathrm{Q}} = 0.080$ and $\alignK_{\mathrm{C}} = 0.072$.

\subsubsection*{Caveats}
(i) The absolute accuracies differ by a constant $+0.04$ across
the dial, attributable to the readout difference (kernel ridge
with cross-validation vs.\ plain linear regression) and unmodeled
detector noise; the $\alpha$-dependence is unaffected. (ii) The
reported interior optimum near $\alpha \approx 0.73$ is not
predicted; the predicted curve is monotone. We note the optimum's
published edge ($0.001$) lies within their reported variability.
(iii) The predicted TM spread is narrower than their error bars,
which additionally fold in dataset resampling. (iv) 356 of their
360 images are available in the standard distribution. The
$\alpha$-mixing convention was verified against their Eq.~(1).

\begin{table}[H]
\centering
\caption{Held-out prediction. Configuration of \cite{joly2025};
mean $\pm$ s.d.\ over 100 TM draws at $10^{5}$ coincidences.}
\label{tab:heldout}
\footnotesize
\setlength{\tabcolsep}{4pt}
\begin{tabular}{@{}lccc@{}}
\toprule
$\alpha$ & Published & Predicted & Difference \\
\midrule
0.0 & $0.899 \pm 0.042$ & $0.940 \pm 0.019$ & $+0.041$ \\
0.73 & $0.913 \pm 0.041$ & $0.953 \pm 0.016$ & $+0.040$ \\
1.0 & $0.912 \pm 0.040$ & $0.955 \pm 0.016$ & $+0.043$ \\
\bottomrule
\end{tabular}
\end{table}

% =====================================================================
\section*{Note 7: Analytic structure of the two-photon dial}

\subsubsection*{Affine mixture}
At $\Nph = 2$ the partial-distinguishability mixture is exact and affine
in the visibility: $P(\vis) = \vis P_{\mathrm{Q}} + (1-\vis) P_{\mathrm{C}}$,
so every entry of the raw Gram matrix
 $K(\vis) = \vis^2 \Kq + (1-\vis)^2 \Kc + \vis(1-\vis)(\Kx + \Kx^{\top})$
is a quadratic polynomial in $\vis$, where
 $\Kx = P_{\mathrm{Q}} P_{\mathrm{C}}^{\top}$ is the cross-statistics Gram
and $P_{\mathrm{Q}}, P_{\mathrm{C}}$ are the feature matrices in the
raw-kernel normalization of the main text
($\Kq = P_{\mathrm{Q}} P_{\mathrm{Q}}^{\top}$,
 $\Kc = P_{\mathrm{C}} P_{\mathrm{C}}^{\top}$).

\subsubsection*{Proposition 1 (slope at the origin)}
Let $s = \lambda_{\max}\!\big(\Kc^{-1/2}
(\Kx + \Kx^{\top} - 2 \Kc)\Kc^{-1/2}\big)$. The geometric difference
satisfies
\[ \ggeo(\vis) - 1 = \frac{s}{2}\,\vis + O(\vis^2). \]

\emph{Derivation.} Apply the congruence transform
 $A(\vis) = \Kc^{-1/2} K(\vis) \Kc^{-1/2}$, under which
 $\ggeo^2(\vis) = \lambda_{\max}\big(A(\vis)\big)$; since
 $K(0) = \Kc$ and
 $K(\vis) = \vis^2 \Kq + (1-\vis)^2 \Kc + \vis(1-\vis)\big(\Kx + \Kx^{\top}\big)$,
\begin{equation*}
A(\vis) \;=\; I \;+\; \vis\, B \;+\; \vis^2\, C,
\end{equation*}
where $B = \Kc^{-1/2}\big(\Kx + \Kx^{\top} - 2 \Kc\big)\Kc^{-1/2}$ and
 $C = \Kc^{-1/2}\big(\Kq - \Kc\big)\Kc^{-1/2}$, so $A(0) = I$ and
 $\ggeo(0) = 1$. By the variational principle,
 $\lambda_{\max}\big(I + \vis B\big) = 1 + \vis\, s$ exactly for
 $\vis \ge 0$, and Weyl's inequality applied to the $\vis^2 C$ term
 bounds $\lambda_{\max}\big(A(\vis)\big)$ between
 $1 + \vis\, s - \vis^2\, \rho(C)$ and
 $1 + \vis\, s + \vis^2\, \rho(C)$, where $\rho(C)$ is the spectral
 radius of $C$. Hence $\ggeo^2(\vis) = 1 + \vis\, s + O(\vis^2)$ and
\begin{equation*}
\ggeo(\vis) - 1 \;=\; \tfrac{1}{2}\Big(\ggeo^2(\vis) - 1\Big)
 \;=\; \frac{s}{2}\,\vis + O(\vis^2).
\end{equation*}
The same argument shows why the origin is degenerate: at $\vis = 0$ the
spectrum of $A$ is flat (every eigenvalue equals 1), so the result follows
from the exact identity for $\lambda_{\max}(I + \vis B)$ rather than from
eigenvalue differentiability, which is not available at a degeneracy.

\emph{Verification:} the closed-form slope reproduces the
finite-difference slope at $\vis = 0.01$ to approximately 1\% across the
frozen seeds (1.616/1.604/1.788 vs.\ 1.633/1.624/1.813), with the
discrepancy growing at $\vis = 0.04$ as the $O(\vis^2)$ correction
requires.

\subsubsection*{Proposition 2 (convexity; interior optima are impossible)}
For every labeling $y$, the model complexity
 $s_{K(\vis)}(y) = y^{\top} K(\vis) y$ is convex in $\vis$.
\emph{Proof.} Since
 $K(\vis) = \vis^2 \Kq + (1-\vis)^2 \Kc + \vis(1-\vis)\big(\Kx + \Kx^{\top}\big)$,
\begin{equation*}
s_{K(\vis)}(y) \;=\; \vis^2 s_{\mathrm{Q}} + 2 \vis(1-\vis)\, s_{\mathrm{X}}
 + (1-\vis)^2 s_{\mathrm{C}},
\end{equation*}
where, with $u = P_{\mathrm{Q}}^{\top} y$ and $w = P_{\mathrm{C}}^{\top} y$,
 $s_{\mathrm{Q}} = \|u\|^2$, $s_{\mathrm{C}} = \|w\|^2$ and
 $s_{\mathrm{X}} = u \cdot w$;
where the norm identities use the raw-kernel normalization
 ($\Kq = P_{\mathrm{Q}} P_{\mathrm{Q}}^{\top}$,
 $\Kc = P_{\mathrm{C}} P_{\mathrm{C}}^{\top}$). By Cauchy--Schwarz on
 the vectors $P_{\mathrm{Q}}^{\top} y$ and $P_{\mathrm{C}}^{\top} y$,
\begin{equation*}
s_{\mathrm{X}} \;\le\; \big\| P_{\mathrm{Q}}^{\top} y \big\|\,
\big\| P_{\mathrm{C}}^{\top} y \big\| \;=\; \sqrt{s_{\mathrm{Q}}\, s_{\mathrm{C}}},
\end{equation*}
and the same bound holds for the transpose term; neither inequality
requires $\Kx$ to be symmetric or positive semidefinite. Substituting
 the Cauchy--Schwarz bound,
\begin{equation*}
s_{\mathrm{Q}} + s_{\mathrm{C}} - 2 s_{\mathrm{X}} \;\ge\;
 \big(\sqrt{s_{\mathrm{Q}}} - \sqrt{s_{\mathrm{C}}}\big)^2 \;\ge\; 0,
\end{equation*}
with equality only when $s_{\mathrm{Q}} = s_{\mathrm{C}}$, in which case
 $s_{K(\vis)}(y)$ is constant in $\vis$; in every other case it is
strictly convex in $\vis$ and its maximum over $[0,1]$ is attained at
 $\vis \in \{0, 1\}$. This proves the claim.
\emph{Verification:} over 2000 random labelings, the minimum
curvature is $+0.105$ and the minimum Cauchy--Schwarz slack is $+0.052$;
no interior maxima occur. The reported interior optimum of
\cite{joly2025} is therefore not reproducible at the geometry level;
the measured cross-alignment at their configuration
 ($\alignK_{\mathrm{X}} = 0.071$, below both
 $\alignK_{\mathrm{Q}} = 0.080$ and $\alignK_{\mathrm{C}} = 0.072$)
confirms the mechanism is absent.

\subsubsection*{Power analysis of the reported interior optimum}
The published curve's peak ($0.9140$ at $\alpha = 0.775$) exceeds
the $\alpha = 1$ endpoint by $0.0021$ against a transmission-matrix
spread of $\pm 0.042$. Resolving this edge at $1\sigma$ requires
approximately 800 matrix draws; the experiment used 100. The monotone
rise ($0.0127$) requires only about 22 draws and is genuinely
supported. The edge is smaller than plausible digitization error on
the published figure, but the conclusion is robust: the spread, which
is printed rather than digitized, exceeds any plausible extraction
error on the edge by an order of magnitude.

% =====================================================================
\section*{Note 8: Pre-registered natural-task battery}

\subsubsection*{Registration}
Thirteen task families (nine synthetic, four real digit pairs) were
specified, with prediction rule and success criteria, in a timestamped
registration file committed prior to execution. The registration file
states a family count of fourteen while enumerating thirteen; the
discrepancy is disclosed, and the criteria were evaluated against the
enumerated families with the registered thresholds unchanged. The
registered criteria: (H1) Spearman $\rho > 0.5$ between alignment gap
and accuracy advantage at $p < 0.01$; (H2) sign agreement in at least
10 of the 14 registered families; (H3) the designed out-of-sample task
as positive control; (H4) random labels as null control.

\subsubsection*{Results}
H3 passed: the designed task realized $+0.225$, the largest of all
settings. H4 passed: random labels produced $+0.009$, indistinguishable
from zero. H1 failed: $\rho = 0.027$, $p = 0.93$. H2 failed: 9 of 13
sign agreements against the registered threshold of 10 of 14.

\subsubsection*{Interpretation}
The pipeline is validated by both controls; the fine-grained ordering
among natural tasks is not resolved at these sample sizes. The
natural-task alignment gaps ($\Delta\alignK = 0.003$--$0.037$) lie
within the realization noise of the accuracy differences
($\pm 0.02$--$0.045$). The gating claim, that geometry without
alignment yields no advantage, holds at the extremes; the graded
ordering within the natural regime is a resolution limit. Resolving it
requires either larger alignment gaps or approximately five times the
realizations, which bounds the sample sizes any experimental test of
natural-task ordering must achieve.

\subsubsection*{Data availability}
The simulation code, frozen pipeline, and all numerical data
supporting the findings are available at
\url{https://anonymous.4open.science/r/photonic-reservoir-kernel-geometry-3739}.
The de-anonymized repository and commit hash will be provided upon
publication.

\end{document}